\documentclass[runningheads]{llncs}

\usepackage{multirow}
\usepackage[T1]{fontenc}
\usepackage{bm}
\usepackage{adjustbox}
\usepackage{graphicx}
\usepackage{footmisc}
\begin{document}
\title{A Neural Denoising Vocoder for Clean Waveform Generation from Noisy Mel-Spectrogram based on Amplitude and Phase Predictions \thanks{This work was funded by the Anhui Provincial Natural Science Foundation under Grant 2308085QF200, the National Nature Science Foundation of China under Grant 62301521 and the Fundamental Research Funds for the Central Universities under Grant WK2100000033.}}
\titlerunning{Neural Denoising Vocoder}
\author{Hui-Peng Du \and Ye-Xin Lu \and Yang Ai\thanks{Corresponding author.} \and Zhen-Hua Ling}
\authorrunning{H.-P. Du et al.}
\institute{National Engineering Research Center of Speech and Language Information Processing, \\University of Science and Technology of China, Hefei, P. R. China \\
	\email{\{redmist, yxlu0102\}@mail.ustc.edu.cn},  \email{\{yangai, zhling\}@ustc.edu.cn}}
\maketitle              % typeset the header of the contribution
\begin{abstract}
This paper proposes a novel neural denoising vocoder that can generate clean speech waveforms from noisy mel-spectrograms. 
The proposed neural denoising vocoder consists of two components, i.e., a spectrum predictor and a enhancement module. 
The spectrum predictor first predicts the noisy amplitude and phase spectra from the input noisy mel-spectrogram, and subsequently the enhancement module recovers the clean amplitude and phase spectrum from noisy ones. 
Finally, clean speech waveforms are reconstructed through inverse short-time Fourier transform (iSTFT). 
All operations are performed at the frame-level spectral domain, with the APNet vocoder and MP-SENet speech enhancement model used as the backbones for the two components, respectively. 
%We utilize the modified APNet as our spectrum predictor, and the MP-SENet as our enhancement module. 
% Experiments show that, despite the lack of phase information and partial amplitude information in the input mel-spectrogram, our proposed method achieves higher PESQ compared to MetricGAN, which takes amplitude spectrum as input.
Experimental results demonstrate that our proposed neural denoising vocoder achieves state-of-the-art performance compared to existing neural vocoders on the VoiceBank+DEMAND dataset.
Additionally, despite the lack of phase information and partial amplitude information in the input mel-spectrogram, the proposed neural denoising vocoder still achieves comparable performance with the serveral advanced speech enhancement methods.
\end{abstract}

\section{Introduction}

A neural vocoder, which converts speech acoustic features to speech waveforms, is an essential component of text-to-speech (TTS) or voice conversion (VC) systems. 
However, current neural vocoders \cite{kong2020hifi,siuzdak2023vocos} mostly consider clean scenarios and lack denoising capabilities. 
In real-world scenarios, speech is often affected by noise. 
In TTS, if we can only obtain noisy speech from a person, and we want to synthesize clean speech for this speaker based on new text, a denoising vocoder which converts the noisy acoustic features to clean waveforms is essential. 
The technology closest to a denoising vocoder is speech enhancement (SE).
SE is a highly practical technology with significant real-world applications, especially for audio devices such as hearing aids and telecommunications. 
Numerous deep learning-based time-frequency (TF) domain \cite{valentini-botinhao2018speech,ai2019dnn} and time-domain SE techniques \cite{pascual2017segan,defossez2020real} have been proposed to improve the perceptual quality and intelligibility of speech, achieving promising results. 
However, these methods typically require knowledge of the complete noisy waveform or at least the noisy amplitude spectrum. 
A denoising vocoder is more challenging than SE because it can only take noisy acoustic features as input, which are often compressed and have lost some information. 
% In real-world scenarios, speech signals are inevitably subject to intrusive noise, leading to distortion of the corresponding acoustic features (e.g. mel-spectrograms). 
% Synthesis systems, such as voice conversion (VC) or text-to-speech (TTS), typically use mel-spectrograms as input features to synthesize waveforms. 
% Therefore, the denoising generation from distorted mel-spectrograms to clean speech waveforms is worth investigating.
% Notably, synthesis systems such as voice conversion (VC) or text-to-speech (TTS) generally utilize mel-spectrograms as intermediate acoustic features.
% In real-world scenarios, it is difficult to avoid interference from devices or channels, making these features impure. 
% Therefore, enhancement techniques for mel-spectrogram are necessary and worth attention.

Although a denoising vocoder is crucial for applications in certain noisy synthesis scenarios, it has not been extensively studied. 
DNR-HiNet \cite{ai2022denoising} achieved clean waveform generation from noisy acoustic features, but it required joint input of fundamental frequency (F0) and mel-spectrogram, which is not compatible with some mainstream TTS and VC methods. 
Therefore, as a preliminary attempt, we propose a neural denoising vocoder to generate clean speech solely from the noisy mel-spectrogram, aiming to assist more real-world applications and hoping to stimulate related research in this area. 
The proposed neural denoising vocoder fully integrates existing vocoding technology and SE technology. 
It features a two-stage pipeline. 
In the first stage, noisy acoustic features undergo a spectrum predictor module, which consists of an amplitude spectrum predictor (ASP) and a phase spectrum predictor (PSP), to predict the noisy amplitude spectrum and noisy phase spectrum, respectively. 
%The vocoder we proposed \footnote{\url{https://redmist328.github.io/denoising-vocoder}} features a two-stage pipeline: In the first stage, noisy acoustic features undergo the spectrum predictor module, which consists of an amplitude spectrum predictor (ASP) and a phase spectrum predictor (ASP), to predict the noisy amplitude spectrum and noisy phase spectrum, respectively. 
Then in the second stage, an enhancement module is employed to predict clean amplitude and phase spectra, which are finally transformed into clean speech waveforms via inverse short-time Fourier transform (iSTFT). 
All operations of this neural denoising vocoder are performed at the frame-level spectral domain, aiming to improve generation efficiency compared to waveform-based methods.
For the spectrum predictor module, inspired by the APNet vocoder \cite{ai2023apnet}, we employed residual convolutional network (ResNet) as the backbone network for both ASP and PSP. 
For the enhancement module, we utilized the state-of-the-art (SOTA) TF-domain SE model MP-SENet \cite{lu2023mp} to accomplish spectral enhancement. 

The rest of this paper is organized as follows. 
In Section \ref{sec:pagestyle}, we give details of our proposed denoising neural vocoder. 
The experimental results and analysis are presented in Section \ref{sec:exp}. 
Finally, we make a conclusion and preview some areas of future research in Section \ref{sec:con}.

\section{Proposed Method}
\label{sec:pagestyle}
\subsection{Overview}
\begin{figure}[t]
	\centering
	\includegraphics[width=\textwidth]{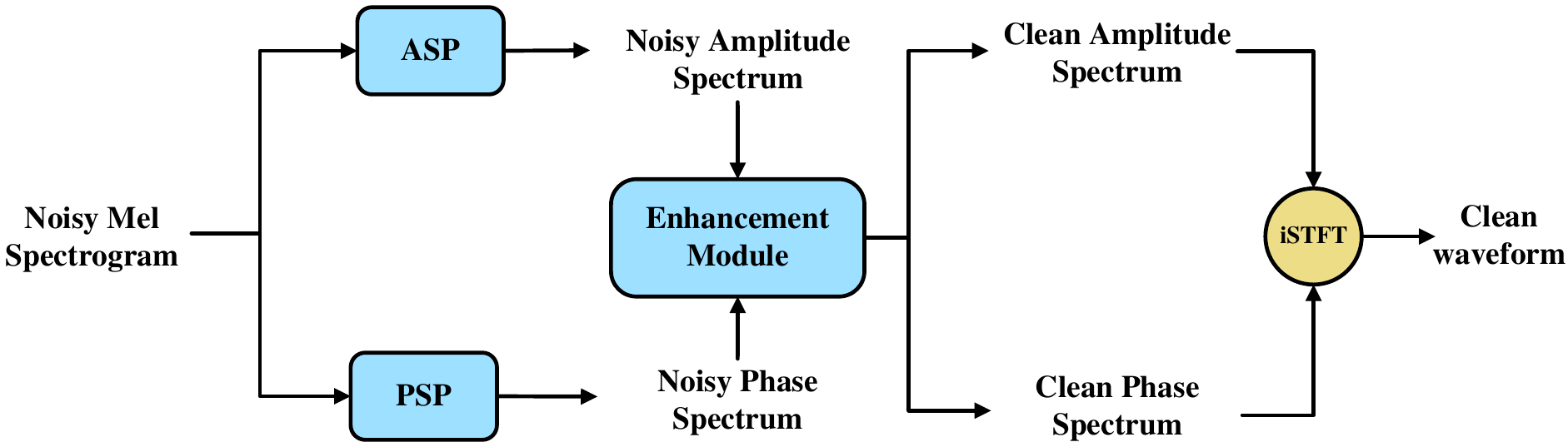}
	\caption{Overall architecture of the proposed neural denoising vocoder. Here, ASP, PSP, and iSTFT stand for amplitude spectrum predictor, phase spectrum predictor, and inverse short-time Fourier transform, respectively.} 
 \label{fig1}
\end{figure}

As depicted in Fig. \ref{fig1}, our proposed neural denoising vocoder pipeline comprises two components. 
In the spectrum predictior, the ASP and PSP forecast noisy amplitude spectrum and phase spectrum from the noisy mel-spectrogram. 
In the enhancement module, the SE model denoises the predicted noisy amplitude and phase spectra to clean ones. 
Finally, the clean speech waveform is reconstructed through iSTFT.

\label{ssec:block}
\subsection{Spectrum Predictor}
\label{subsec: spectrum}

Firstly, we provide evidence for the necessity of introducing a spectral predictor through some analysis. 
Alternatively, the noisy mel-spectrogram could be directly passed through an enhancement model to generate clean speech waveforms. 
However, we found this approach to be infeasible. 
% For SE models based on amplitude and phase spectrum enhancement, the performance is poor if the noisy input lacks phase information. 
% Therefore, The phase spectrum of noisy speech is essential for SE.
%Following, we provide some evidences for this.
With $\bm{N}_a$ and $\bm{N}_p$ respectively denoting the amplitude and phase spectrum of the additive noise signal, the noisy phase $\bm{Y}_p$ of the noisy speech can be calculated by the following equation, i.e.,
\begin{equation}
    \bm{Y}_p=Angle\left(\bm{X}_a e^{j\bm{X}_p}+\bm{N}_a e^{j\bm{N}_p}\right)
    =\bm{X}_p+Angle\left(1+\frac{\bm{N}_a}{\bm{X}_a}e^{j(\bm{N}_p-\bm{X}_p)}\right),
    \label{eq1}
\end{equation}
where $\bm{X}_a$ and $\bm{X}_p$ denote the clean amplitude and phase spectra, respectively. 
In the case of a high signal-to-noise ratio (SNR), i.e., $\bm{X}_a$ is much larger than $\bm{N}_a$, the second item in Eq. \ref{eq1} would approach zero, which means the noisy phase approximates the clean one.
However, as the SNR decreases, the noisy phase increasingly deviates from the clean phase, resulting in a significant impact on speech perceptual quality and intelligibility. 
Therefore, the recovery of phase information is particularly important in low SNR scenarios. 
However, in preliminary experiments, we found that directly recovering clean phase information from noisy mel-spectrograms using SE model is very challenging, especially in low SNR scenarios. 
This indicates that without prior noisy phase information, it is difficult to directly and accurately recover the important clean phase spectrum from the mel-spectrogram.

%In preliminary experiments, we found that directly recovering clean phase information from noisy mel-spectrograms is very challenging. 
Therefore, we adopt a two-stage prediction approach using noisy amplitude and phase spectra as bridged features. 
To predict them from the noisy mel-spectrogram, inspired by the APNet vocoder \cite{ai2023apnet}, the spectrum predictor is composed of an ASP and a PSP. 
Specifically, APNet can explicitly predict amplitude and phase spectra in parallel with the aid of phase anti-wrapping losses through all-frame-level convolutions. 
APNet2 \cite{du2023apnet2} has made improvements over APNet, and we have adapted some of these enhancements for our spectrum predictor. 
Specifically, we incorporate a multi-resolution discriminator (MRD) instead of the previous multi-scale discriminator (MSD) to constrain waveform generation across multiple resolutions for the adversarial training. 
Additionally, we replace the cosine-shaped anti-wrapping losses with linear ones \cite{ai2023neural} to ensure gradient consistency across all phase values. 
Our preliminary attempts revealed that the ConvNeXt v2 \cite{woo2023convnext} blocks used in APNet2 were not suitable for our current task, so we retained the ResNet blocks as the backbone of ASP and PSP. 
The predicted noisy amplitude and phase spectrum are reconstructed into a noisy speech waveform through iSTFT and then denoised in the subsequent module.

\subsection{Enhancement module}

The enhancement module first re-extracts the noisy amplitude and phase spectra from the noisy speech predicted by the previous module and then denoises them.
MP-SENet \cite{lu2023mp} is currently one of the SOTA model in SE filed, so we use it to remove the noise from the amplitude spectrum and phase spectrum provided by the spectrum predictor, within the enhancement module. 
It first utilizes Conformer blocks for feature extraction from noisy amplitude and phase spectra. 
Subsequently, an amplitude spectrum mask decoder and a phase decoder are employed to predict mask values for the noisy amplitude spectrum and the predicted clean phase. 
Finally, it reconstructs the clean speech waveform through iSTFT.
For the training of the MP-SENet, we follow the methodology outlined in \cite{lu2023mp}. 
Specifically, we employ loss functions defined on the time domain, amplitude spectrum, and complex spectrum, as well as anti-wrapping losses based on the phase spectrum \cite{ai2023neural}. 
Additionally, we utilize generative adversarial training as suggested by \cite{fu2019metricgan} during the training process.

\label{ssec:loss}

\section{Experiments}
\label{sec:exp}

\subsection{Experimental Setup}
\label{ssec:expset}
\subsubsection{Dataset.} 
In our experiments, we utilized the VoiceBank+DEMAND dataset \cite{valentini-botinhao2016investigating}, comprising paired clean and noisy audio clips sampled at 48 kHz. 
The clean speech subset, sourced from the Voice Bank corpus \cite{veaux2013voice}, consists of 11,572 clips from 28 speakers for training, and 872 clips from 2 speakers for testing. 
Each clean clip was mixed with 10 types of noise (8 from the DEMAND database \cite{thiemann2013diverse} and 2 artificial types) at SNRs of 0 dB, 5 dB, 10 dB, and 15 dB for training, and 5 types of unseen noise from the DEMAND database at SNRs of 2.5 dB, 7.5 dB, 12.5 dB, and 17.5 dB for testing. 
All speech clips were resampled to 16 kHz for consistency in the experiments.

\subsubsection{Implementation.}\label{above}
In the proposed neural denoising vocoder\footnote{Generated speech can be found at \url{https://redmist328.github.io/denoising-vocoder}.}, for both the ASP and PSP, we utilized 80-dimensional mel-spectrograms with a frame shift of 80 (5 ms), frame length of 320 (20 ms), and 1024-point FFT as input. 
For the enhancement module, we use the open source checkpoint of MP-SENet\footnote{\url{https://github.com/yxlu-0102/MP-SENet}.}. 
Input features of the enhancement module were extracted from waveforms with FFT point, frame length, and frame shift set to 400, 400 (25 ms), and 100 (6.25 ms), respectively. 
Training of all models employed the AdamW optimizer \cite{loshchilov2018decoupled} for 700k steps, with the learning rate initialized at 0.0002. 
A detailed discussion on the implications of these differences in amplitude and phase spectrum configurations on the results of the SE models will be provided in Sec. \ref{sssec:ee}.

\subsubsection{Baselines.} 
For comparison with neural vocoders, we evaluated two renowned vocoders, i.e., HiFi-GAN\footnote{\url{https://github.com/jik876/hifi-gan}.} \cite{kong2020hifi} and Vocos\footnote{\url{https://github.com/gemelo-ai/vocos}.} \cite{siuzdak2023vocos}, which directly use noisy mel-spectrogram as input and outputs the clean speech waveforms. 
Both HiFi-GAN and Vocos were trained following their original methodologies, keeping the hyperparameters for feature extraction consistent with ours. 
Regarding the comparison of SE models, we utilized the results of MetricGAN \cite{fu2019metricgan} and MetricGAN+ \cite{fu2021metricgan+}, which only used amplitude spectra as inputs.

\subsubsection{Evaluation metircs.}

To assess the enhanced speech quality, we selected six commonly used objective evaluation metrics, including perceptual evaluation of speech quality (PESQ), segmental signal-to-noise ratio (SSNR), short-time objective intelligibility (STOI), and three composite measures (i.e., CSIG, CBAK, and COVL). 
The composite measures predict the mean opinion score (MOS) for signal distortion, background noise intrusiveness, and overall effect, respectively. 
Higher values across all these metrics indicate better performance.

\label{ssec:eva}
\subsection{Comparison with Advanced Neural Vocoders}
We first compared our proposed neural denoising vocoder with two advanced neural vocoders (i.e., HiFi-GAN and Vocos) and the objective results are depicted in Table~\ref{tab1}. 
Both HiFi-GAN and Vocos are directly trained to predict clean speech waveforms from the input noisy mel-spectrograms. 
Overall, our proposed neural denoising vocoder significantly outperformed both HiFi-GAN and Vocos in all metrics.
Since both the baseline vocoders were directly adopted for the denoising task without modification, their limited performance indicated the difficulty of directly recovering clean waveforms from noisy mel-spectrograms.
This is probably because SE highly requires global information, while current vocoders, which are mostly based on convolutional neural networks (CNNs), primarily focus on short-term and local information. 
Additionally, compared to SE methods, the input to the denoising vocoder cannot contain complete speech information, making the modeling more challenging. 
Therefore, without any further processing, current vocoders are unable to achieve effective denoising.
Nevertheless, our proposed neural denoising vocoder’s enhancement module used multi-layer, two-stage Conformer (TS-Conformer) blocks that alternately capture both local and global information in the time and frequency domains, achieving better enhancement performance.

\begin{table}[t]
\renewcommand{\arraystretch}{1.1}
	\centering
	\caption{Objective experimental results of the proposed neural denoising vocoder and several advanced neural vocoders on the VoiceBank+DEMAND test set.}\label{tab1}
	%\adjustbox{width=0.7\textwidth}{
		\begin{tabular}{l c c c c c c}
			\hline
			\hline
			 & {PESQ}& {CSIG} & {CBAK} & {COVL} & {SSNR} & {STOI}\\
			\hline
			{Noisy} &1.97&3.35&2.44&2.63 &\textbf{ {1.68}}&0.91\\
			\hline
       			{HiFi-GAN} & 2.24&3.97&2.54&3.14&-1.37 &0.92\\ 
  			{Vocos} & 1.45&2.80&2.09&2.14&-2.14 &0.80\\ 
		{Proposed} &\textbf{2.88}&\textbf{4.41}&\textbf{3.06}&\textbf{3.69}&1.45&\textbf{0.94}\\
			%\hline
			
			\hline
			\hline
	\end{tabular}%}
\end{table}

\vspace{-1.5mm}
\subsection{Comparison with advanced SE methods}
\vspace{-1.5mm}
We further compared our proposed neural denoising vocoder with two SE models (i.e., MetricGAN and MetricGAN+) and the objective results are depicted in Table~\ref{tab2}.
Notably, the results for SSNR and STOI were not provided in the original paper of MetricGAN \cite{fu2019metricgan} and MetricGAN+ \cite{fu2021metricgan+}. 
Therefore, the results of these two metrics are excluded in Table~\ref{tab2}. 
It can be observed that our proposed method outperformed MetricGAN across PESQ, CSIG, and COVL, but still fell in CBAK. 
As for the comparison with MeticGAN+, our proposed neural denoising vocoder generated speech with better CSIG and COVL, but performed worse in terms of PESQ and CBAK.
This performance gap was reasonable, as both of these SE models take amplitude spectrum as input, whereas our proposed method utilized mel-spectrogram, which lacked some fine-grained frequency information.
Overall, our proposed denoising vocoder was comparable in performance to these advanced SE methods, but there is still room for improvement. 

\begin{table}[t!]
\renewcommand{\arraystretch}{1}
	\centering
	\caption{Objective experimental results of the proposed neural denoising vocoder and several advanced SE methods on the VoiceBank+DEMAND test set.}\label{tab2}
	%\adjustbox{width=0.7\textwidth}{
		\begin{tabular}{l c c c c c }
			\hline
			\hline
			 &Input& {PESQ}& {CSIG} &{CBAK}& {COVL}\\
			\hline
			{Noisy} & -&1.97 & 3.35 & 2.44 & 2.63 \\
            \hline
			%\hline
  			{MetricGAN} & Amplitude&2.86&3.99&\textbf{3.18}&3.42 \\ 
       		{MetricGAN+} & Amplitude&\textbf{3.15}&4.14&3.16&3.64 \\ 
			{Proposed} & Mel-spectrogram&2.88&\textbf{4.41}&3.06 &\textbf{3.69}\\	
			\hline
			\hline
	\end{tabular}%}
\end{table}

\begin{table}[t]
\renewcommand{\arraystretch}{1.2}
	\centering
	\caption{Objective experimental results for analysis experiments on the VoiceBank+DEMAND test set.}\label{tab3}
	%\adjustbox{width=0.75\textwidth}{
		\begin{tabular}{l c c c c c c }
			\hline
			\hline
			\multirow{1}{*} & {PESQ}& {CSIG} &{CBAK}& {COVL} &{SSNR}& {STOI}\\
			\hline
			{Proposed} & \textbf{2.88}&\textbf{4.41}&\textbf{3.06}&\textbf{3.69}&\textbf{1.45} & \textbf{0.94}\\
			%\hline
  			%{APNet*} & 2.35&3.97&2.72&3.20&0.07 &0.93\\ 
          {Proposed rep. APNet} & 2.76&4.30&2.97&3.58&0.96&\textbf{0.94}\\
          {Proposed w/ matching STFT condition} & 2.26&3.91&2.54&3.14&--0.31 &0.91\\ 
			%\hline
			
			\hline
			\hline
	\end{tabular}%}
\end{table}
\subsection{Analysis and Discussion}
\label{sssec:ee}

As mentioned in Section \ref{subsec: spectrum}, the spectrum predictor is modified from APNet. 
Compared to the original APNet \cite{ai2023apnet}, the spectrum predictor has made some structural and loss improvements. 
To verify the effectiveness of this improvement, we directly replaced the spectrum predictor in the proposed neural denoising vocoder with the original APNet. 
The experimental results are shown in the second-to-last row of Table~\ref{tab3}. 
We can see that the proposed neural denosing vocoder with spectrum predictor performed superior compared to the one with original APNet, suggesting that the modifications to the APNet were quite effective and applicable to denoising tasks. 

Subsequently, we discuss issues related to STFT configuration. 
Due to the inconsistency between the feature extraction parameters used by the spectrum predictor and the enhancement module, in our implementation, we used the iSTFT to recover the noisy amplitude and phase spectra outputted by the spectrum predictor into noisy speech waveform. 
Subsequently, the recovered noisy speech waveform underwent STFT with different feature extraction parameters, and the extracted spectra were then fed into the enhancement module for denoising.
To investigate the impact of parameter inconsistency, we used the STFT configuration that matches MP-SENet for the spectrum predictor, i.e., 400-point FFT, 400-point frame length, and 100-point frame shift. 
The experimental results are shown in the last row of Table~\ref{tab3}. 
It can be seen that deliberately matching the feature extraction parameters during direct training of the two stages has a negative impact on the model performance, indicating that each stage should still be trained with its most suitable parameters. 

% compared our proposed method with the original APNet under the condition of two feature extraction parameters, i.e., the original feature extraction parameters (1024-point FFT, 320-point frame length, and 80-point frameshift) and the feature extraction parameters matching those of the enhancement model MP-SENet (400-point FFT, 400-point frame length, and 100-point frameshift). 
% The same MP-SENet model was used for the enhancement in both cases. 
% The experimental results are illustrated in Table~\ref{tab3} (* denotes the matching condition in the experiment).  
% It can be seen that deliberately matching the feature extraction parameters during direct training of the two stages has a negative impact on the model performance, indicating that each stage should still be trained with its most suitable parameters. 
% Furthermore, the proposed modified APNet performed superior compared to the original APNet under the original feature extraction parameters, suggesting that the modifications to the APNet were quite effective.
% we found that our modifications to the APNet were quite effective , suggesting that the modifications to the methods during the design of each stage module were more effective for improving performance than adjusting the feature extraction parameters.

\section{Conclusions}
This paper presents a novel neural denoising vocoder that is capable of converting input noisy mel-spectrogram into clean speech. 
This vocoder consists of a spectrum predictor and an enhancement module, combining the functionalities of both vocoding and denoising.
The spectrum predictor module estimates the noisy amplitude and phase spectra from the input noisy mel-spectrogram, while the enhancement module refines these noisy spectra to obtain the clean ones. 
Subsequently, the clean speech is synthesized using the iSTFT. 
%In our approach, we employ a modified APNet as the spectrum predictor and the MP-SENet as the enhancement module. 
Experimental results demonstrate that, despite the absence of phase information and partial amplitude information in the input mel-spectrogram, our proposed neural denoising vocoder still outperforms baseline vocoders and is comparable to several SE methods. 
Further exploration of building an end-to-end denoising vocoder without the need for a noisy speech bridge will be our future work.
\label{sec:con}

\bibliographystyle{splncs04_unsort}
\bibliography{mybibliography}

\end{document}